# Hotel Preference Rank based on Online Customer Review

Muhammad Apriandito Arya Saputra[1], Andry Alamsyah[2], Fajar Ibnu Fatihan[3]
School of Economics and Business
Telkom University
Bandung, Indonesia
[1]apriandito@student.telkomuniversity.ac.id, [2]andrya@telkomuniversity.ac.id, [3]fajaribnufatihan@gmail.com

*Abstract* — Top line hotels are now shifting into the digital way in how they understand their customers to maintain and ensuring satisfaction. Rather than the conventional way which uses written review or interviews, the hotel is now heavily investing in Artificial Intelligence particularly Machine Learning solutions. Analysis of online customer reviews changes the way companies make decisions in a more effective way than using conventional analysis. The purpose of this research is to measure hotel service quality. The proposed approach emphasizes on service quality dimensions reviews of the top-5 luxury hotel in Indonesia that appear on online travel site TripAdvisor based on section "Best of 2018". In this research, we use a model based on a simple Bayesian classifier to classify each of customer review into one of service quality dimensions. Our model was able to separate each classification properly by accuracy, kappa, recall, precision, and F-measure measurements. To uncover latent topics in the customer opinion we use Topic Modelling. We found that the common issues occur is about responsiveness as it got the lowest percentage compared to others. Our research provides a faster outlook of hotel rank based on service quality to end customer based on a summary of the previous online review.

## I. INTRODUCTION

Data generated by customers in form of reviews spread across the web have become an essential in keeping relation with your customers. There is massive flow of data spread across the internet containing valuable information about customers if it is done right. Every little knowledge a company knows about their customers can increase the opportunity to obtain more revenue. One type of data grows at significant rate is data generated by the user itself called user-generated content.

Digital customers-to-customers interactions in various social media platforms has enabled the fast growth of user-generated content. User-generated content (UGC), in the form of product reviews and ratings, shows a huge potential to reduce information asymmetries in tourism markets, related to product quality and customer segment specific suitability [1]. It can be used as a source of information related to what customer think to improve operation in the hospitality industry. Customer feedback is an essential source of information for improving operations in the service industry [2]. But to capture that valuable information including accurate, complete picture, and fast time processing are always been a challenge.

In the hospitality industry, planning, comparing and booking accommodation are important processes. Customers' decision about their accommodation are often influenced by the past experience of other customers. Customers usually get advice from their relations such as colleagues, family member, or friends. This is one kind of viral marketing called word-of-mouth (WOM). WOM is a powerful marketing tool [3]. Information obtained via word-of-mouth are considered more reliable and trusted among customers because they are getting the actual experiences - advantages and disadvantages - of people who have been there rather than reading an advertisement. Now, as the rise of social media, the way of word-of-mouth spread is shifting into the digital way.

Online reviews and rating in the hospitality industry gained a reputation as it becomes one of the factors in the average daily rate of a hotel. If a hotel increases its review scores by 1 point on a 5-point scale e.g., from 3.3 to 4.3, the hotel can increase its price by 11.2 percent and still maintain the same occupancy or market share [4]. The reason for this is because the social media has the potential to move markets just by driving consumers' purchasing patterns.

Top line hotels are now shifting into the digital way in how they understand their customers to maintain and ensuring the satisfaction. Rather than the conventional way which uses written review or interviews, the hotel is now heavily investing in Artificial Intelligence particularly Machine Learning solutions [14]. Machine learning is a method that related to automated large-scale data analysis [13]. Customers' text analysis is changing the way the company making a decision because it provides a more effective way rather than conventional analysis. There are several algorithms for analysis including K-Nearest Neighbours (KNN), Naive Bayes Classifier (NBC), Support Vector Machine (SVM), and Artificial Neural Network (ANN).

In this research, we use Naïve-Bayes Classifier to calculate accuracies, precisions, and recall values of the dimension of SERVQUAL which is based from service quality dimension [5]. Service quality is a conceptual model to assess services based on its performance's percept [6]. Naïve Bayes Classifier is used because of its simplicity in both training and classifying stage [7]. Naïve Bayes method allows each attribute towards the final decision equally and independently from the other attributes. Hence, it is more efficient and more accurate compared to other classifiers [8]. Then to uncover and interpreting the topic we use Topic Modelling. It enables deep inspection of topic-term relationships in an LDA model, while simultaneously providing a global view of the topics, via their prevalence and similarities to each other, in a compact space [9].

The proposed approach emphasizes on SERVQUAL reviews of top-5 luxury hotels in Indonesia that appear on an online travel site called www.tripadvisor.com. The top-5 luxury hotel in Indonesia is based on section "Best of 2018" in TripAdvisor website.

The objectives of this research are first to mapping all reviews to SERVQUAL and second to summarize opinion. The opinion found within reviews will provide useful information for many different purposes and can be categorized into several dimensions. By getting a clearer picture of what customers' thought, hotels may identify areas where customers are satisfied or dissatisfied, and devise strategies for enhancing the quality of service and gain more revenue.

II. LITERATURE REVIEW

*A. Text Analytics*

A term that describes technology and process both, a mechanism for knowledge discovery applied to documents, a means of finding value in text. Solutions mine documents and other forms of 'unstructured' data. It analyses the linguistic structure and applies statistical and machine-learning techniques to discern entities such as names, dates, places, terms, and their attributes as well as relationships, concepts, and even sentiments. It extracts these 'features' to databases for further analysis and automates classification and processing of source documents. It exploits visualization for exploratory analysis of discovered information.

*B. Naïve-Bayes Classifier*

Naïve-Bayes is a simple Bayesian classifier. Naïve-Bayes classifiers assume that the effect of an attribute value on a given class is independent of the values of the other attributes. This assumption is called class-conditional independence. It is made to simplify the computations involved and, in this sense, is considered "naïve" [10]. The advantage of this algorithm is that it only requires a small amount of training data to predict the means and variances of the variables for classification [11].

*C. Text classification performance evaluation*

The performance evaluation was conducted to test the results of classification by measure the performance of the system that has been created. The measurement can be done by confusion matrix method as follow:

TABLE I.　CONFUSSION MATRIX

|  |  | Actual | |
|---|---|---|---|
|  |  | *Success* | *Not Success* |
| **Predicted** | *Success* | True Positive (TP) | False Negative (FN) |
|  | *Not Success* | False Positive (FP) | True Negative (TN) |

- TP (True Positives): class predicted by the predicted positive and negative grade classification system.
- TN (True Negatives): class predictable and predicted negative by a negative grade classification system.
- FP (False Positives): class predicted by the negative but positive grade classification system.
- FN (False Negatives): positive grade but predicted by negative grade classification system.

Mathematical equations and explanations of the above parameters are described below:

- Recall: this measurement to identify how many methods can remember with the correct grouping of text.

$$Recall = \frac{TP}{TP + FN} \quad (1)$$

- Precision: the ratio of measurement on how appropriate the method of textual classification in predicting text.

$$Precision = \frac{TP}{TP + FP} \quad (2)$$

- Accuracy: in order to test the accuracy of the text classification method that has been made.

$$Accuracy = \frac{TP + TN}{TP + FP + TN + FN} \quad (3)$$

- F-Measure: F-Measure is a combination of measurements of precision and recall.

$$F - Measure = 2 \times \frac{Precision \times Recall}{Precision + Recall} \quad (4)$$

- Kappa: Kappa is used to measure agreement between each pair of annotators in which the annotator is used to assess the making of a text classification method.

$$K = \frac{P(A) - P(E)}{1 - P(E)} \quad (5)$$

Classification methods that show the quality and performance of text classification methods that have been constructed by researchers. If the value of kappa is more than 0.75 then the text classification method can be concluded very well [17].

TABLE II.　AGREEMENT MEASURE FOR CATEGORICAL DATA

| **Kappa Score** | **Description** |
|---|---|
| < 0.00 | Poor |
| 0.00 - 0.20 | Slight |
| 0.021 - 0.40 | Fair |
| 0.41 - 0.60 | Moderate |
| 0.61 - 0.80 | Substantial |
| 0.81 - 1.00 | Almost Perfect |

*D. Topic modelling*

A topic model is a statistical model used to identify the latent topics that occur in a collection of text documents; topic modelling is therefore often considered to be a text mining tool for the discovery of hidden semantic structures in a text body [12].

As in this research, we use the simplest topic model which is LDA - it helps to identify the topical features of documents by postulating that documents are described by a topic distribution and that each topic is made up of a distribution of words [8]. The idea behind LDA is to model documents as arising from multiple topics, where a topic is defined to be a distribution over a fixed vocabulary of terms [13]. LDA casts this intuition into a hidden variable model of documents. Hidden variable models are structured distributions in which observed data interact with hidden random variables. With a hidden variable model, the practitioner posits a hidden structure in the observed data and then learns that structure using posterior probabilistic inference [13].

## III. METHODOLOGY

The research stages are divided into 4. Each stages describe activity in doing the research. Data collection talks about activity in collecting the data. Second stage is pre-processing data which is preparing the data to be analysed. Data classification talks about data classifying. Topic modelling talks about visualizing the topic of data in each document.

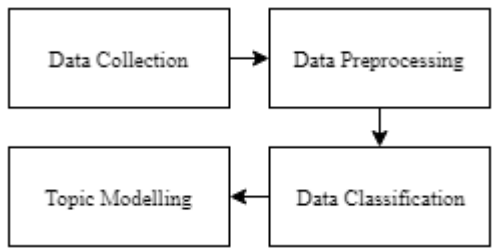

Fig. 1. Research Stages

### A. Data Collection

The data is crawled from TripAdvisor website. The data period is from the very first review until February 2018 with total number of data collected 7,259 reviews.

### B. Data Pre-processing

Pre-processing activities through 3 processes namely, tokenization, filtering and stemming. Tokenize serves to remove all non-letter characters and to divide a text into small, meaningful elements such as sentences and words. Stem is a process to convert the word to its original word. Filter stop words serves to eliminate unnecessary words in tweet. In addition, there is a determination of weighting options, and for this study used TF-IDF.

### C. Data Classification using NBC

The data are divided into 2, training data and testing data. The ratio for each is 30%:70% [7]. The data extracted randomly to build the training dataset for the classifier and the rest is for test data. We use 5 training data and each hotel is different. The categorization of each sentence to SERVQUAL dimensioning can be seen in table 3.

TABLE III. CATEGORY CLASSIFICATION

| Text | Dimension |
|---|---|
| The staffs are beyond brilliant, each one genuinely Lovely and so helpful. | Assurance |
| Staffs greet by name. | Empathy |
| Very attentive and pro-active staffs | Responsiveness |
| This is an amazing hotel in the jungle with beautiful view. | Tangible |

### D. Topic Modelling

Data used for topic modelling is divided into each hotel. The tool used is pyLDAvis package in Python language.

## IV. RESULTS

### A. Classification Results

Accuracy, kappa, precision, recall and F-measure are methods used for evaluating the performance of sentiment analysis. The performance evaluation results of category are followed.

TABLE IV. MODEL EVALUATION RESULT

| Hotel | Accuracy | Kappa | Precision | Recall | F-Measure |
|---|---|---|---|---|---|
| Mandapa, A Ritz Carlton Reserve | 84.09% | 0.786 | 86.06% | 85.46% | 85.76% |
| Komaneka at Tanggayuda | 73.81% | 0.646 | 71.22% | 73.52% | 72.35% |
| Viceroy | 73.33% | 0.639 | 77.77% | 72.39% | 74.98% |
| Katamama | 87.65% | 0.835 | 89.43% | 87.46% | 88.52% |
| Jamahal, Private Resort and Spa | 86.41% | 0.814 | 86.37% | 85.62% | 86.03% |

Based on Table 4 above, it shows that the results of sentiment classification using NBC have a good accuracy percentage. This proved by the kappa score of each company generated by NBC stands between 0.61-0.80 which classified as substantial agreement. The F-measure from each hotel is showing the harmonization of precision and recall values from the classification process works well. The interaction that related to each hotel is mapped to understand the classification. The table 5 below shows the classification percentage of each category.

TABLE V. CLASSIFICATION RESULT BASED ON CATEGORY

| Hotel | Assurance | Empathy | Responsive | Tangibles |
|---|---|---|---|---|
| Mandapa, A Ritz– Carlton Reserve | 49% | 16% | 5% | 30% |
| Komaneka at anggayuda | 35% | 8% | 10% | 47% |
| Viceroy | 7% | 13% | 24% | 56% |
| Katamama | 39% | 11% | 4% | 46% |
| Jamahal, Private Resort and Spa | 26% | 32% | 4% | 38% |

## B. Topic Modelling Result

The blue bar width represents the broad frequency of each term, and the width of the red bar represents the specific frequency of the topic for each term. There is a slider in each of the picture (in the tools) in which allow users to hover the λ which can alter the rankings of terms to help to interpret the topic. By comparing the widths of the red and blue bars for a given term, users can quickly understand whether a term is highly relevant to the selected topic because of its lift (a high ratio of red to blue), or its probability (absolute width of red) [9].

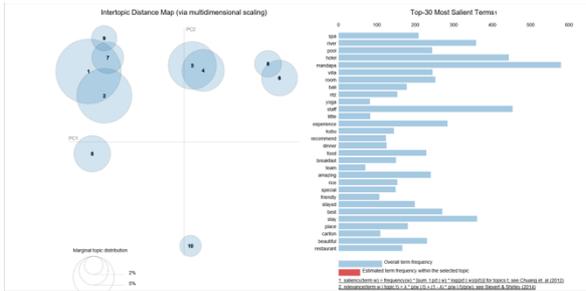

Fig. 2. Mandapa Hotel Topic Modelling Result

In figure 2 for the Mandapa Hotel, based on the top-3 words the topic mainly talks about "the service by the staff".

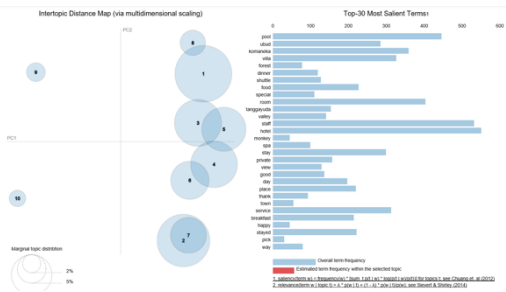

Fig. 3. Komaneka Hotel Topic Modelling Result

For the Komaneka Hotel, in figure 3 the topic mainly talks about "the pool in the villa".

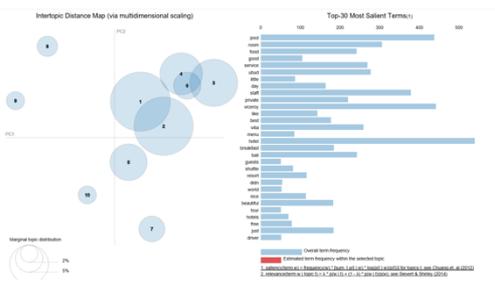

Fig. 4. Viceroy Hotel Topic Modelling Result

In figure 4, for Viceroy Hotel, topic mainly talk about "the breakfast at the restaurant and the spa".

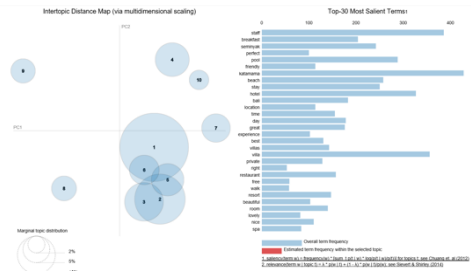

Fig. 5. Katamama Hotel Topic Modelling Result

In figure 5, for Katamama Hotel, the main topic talk about "the staff breakfast and Seminyak".

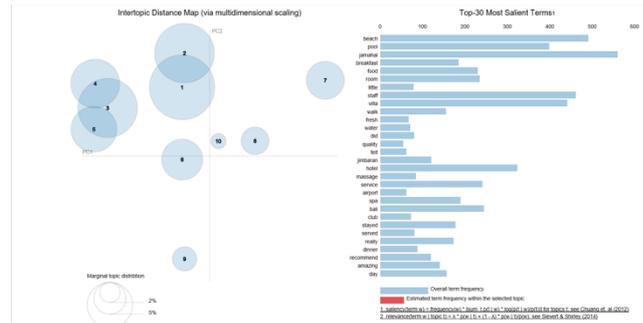

Fig. 6. Jamahal Hotel Topic Modelling Result

And the last, in figure 6. The main topic about Jamahal Hotel is "the beach pool and jamahal".

## V. CONCLUSION

In this research, data classification is capable to separate each service quality dimension properly. It is proven by the percentage of Accuracy, Kappa, Recall, Precision, and F-Measure. To uncover latent topics in the opinion author employ topic modelling. We found in each hotel, the common issues occur is about responsiveness as this dimension got the lowest percentage compared to other dimensions. For each hotel the issues are varied. Mandapa got the lowest percentage for responsiveness. Komaneka is low in the empathy. For Viceroy the assurance needs to be improved. Katamama and Jamahal got the same issue in which low in responsiveness as well. After knowing the lowest dimension of each hotel, we can relate to the topic mostly discuss in each hotel from the topic of review that we have uncovered.

REFERENCES

[1] G. Eason, B. Noble, and I. N. Sneddon, "On certain integrals of Lipschitz-Hankel type involving products of Bessel functions," Phil. Trans. Roy. Soc. London, vol. A247, pp. 529–551, April 1955. *(references)*

[2] Schmunk. S., Hopken. W., Fuchs. M., and Lexhagen M, "Sentiment Analysis: Extracting Decision-Relevant Knowledge from UGC" Information and Communication Technologies in Tourism, Proceedings of the International Conference, 2014.

[3] Han. J. H., Mankad. S., Gavirneni. S., Verma. R, What Guests Really Think of Your Hotel: Text Analytics of Online Customer, Reviews CHR Reports, 2016.

[4] Kotler. P., Keller K.L. Marketing Management 15th Edition New York (Pearson Education), 2016, pp. 645-64.

[5] Anderson. C, The Impact of Social Media on Lodging Performance CHR Reports, 2012.

[6] Parasuraman. A., Zeithaml A. V., Berry. L. L. SERVQUAL: A Multiple Item Scale for Measuring Customer Perceptions of Service Quality Journal of Retailing 64, 1988, pp. 12-40.

[7] Alamsyah A., Rachmadiansyah I. "Mapping online transportation service quality and multiclass classification problem solving priorities", International Conference on Data and Information Science Journal of Physics: Conference Series 971, 2017

[8] Ting. L. S., Ip W. H., Tsang C. H. A. "Is Naïve Bayes a Good Classifier for Document Classification?" International Journal of Software Engineering and its Applications, 2011, pp. 37-45.

[9] Xhemali. P., Hinde. J. C., Stone. G. R. Naïve Bayes vs. Decision Trees vs. Neural Networks in the Classification of Training Web Pages IJCSI, 2009.

[10] Sievert. C., Shirley E. K., LDAvis: A method for visualizing and interpreting topics Proceedings of the Workshop on Interactive Language Learning, Visualization, and Interfaces, 2014, pp. 69-70.

[11] Han. J., Kamber M., Pei J, Data Mining Concepts and Techniques 3rd Edition MA Elsevier Inc, 2012, p. 350.

[12] Syarif. I. Comprehensive Review of Classification Algorithms for High Dimensional Datasets ePrints Soton, 2014, p. 16.

[13] Ko N., Jeong B., Choi S., Yoon J., Identifying Product Opportunities Using Social Media Mining: Application of Topic Modelling and Chance Discovery Theory Volume IEEE 6, 2017, p. 1681

[14] Blei D. M., Lafferty J. D. Topic Models Text Mining: Classification, Clustering, and Applications vol 10 New York CRC Press 34, 2009.

[15] Arusada N. D. M., Putri S. A. N., Alamsyah." A Training Data Optimization Strategy for Multiclass Text Classification" The 6th ICoICT, 2017.

[16] Alamsyah. A., Ginting M. D. Analyzing Employee Voice Using Real-Time Feedback The 4thh UGM ICST, 2018.

[17] Davis J., Goadrich M. The relationship between Precision-Recall and ROC curves Proceedings of the 23rd International Conference on Machine Learning ICML '06, 2006.

[18] Landis R. J., Koch G. G The Measurement of Observer Agreement for Categorical Data Biometrics vol 33, 1977.